\documentclass[preprint,12pt]{elsarticle}

\usepackage{multirow}
\usepackage{pifont}
\usepackage{amssymb}
\usepackage{rotating}
\usepackage{adjustbox}
\usepackage{amsmath}
\usepackage{url}
\usepackage{tabularx}
\usepackage{hyperref}
\usepackage{color,float}
\usepackage{setspace}
\usepackage{soul}
\usepackage{booktabs}
\usepackage[colorinlistoftodos]{todonotes}
\journal{Nuclear Instruments and Methods A}

\begin{document}

\begin{frontmatter}



\title{An Online Data Analysis Framework for Small-Scale Physics Experiments} 

\author[oxford]{H. Ramm\corref{cor1}}
\ead{hayden.ramm@physics.ox.ac.uk}
\cortext[cor1]{Corresponding author}

\author[gsi]{P. Simon}
\author[liverpool,cern]{P. Alexaki}
\author[lancaster]{C. Arran}
\author[ral]{R. Bingham}
\author[cern]{A. Goillot}
\author[iceland,kth]{J. T. Gudmundsson}
\author[ral]{J. W. D. Halliday}
\author[oxford]{B. Lloyd}
\author[oxford]{E. E. Los}
\author[oxford,cern]{V. Stergiou}
\author[oxford]{S. Zhang}
\author[]{G. Gregori\textsuperscript{a} and N. Charitonidis\textsuperscript{d}}

\affiliation[oxford]{organization={Department of Physics, University of Oxford},
            addressline={Parks Road}, 
            city={Oxford},
            postcode={OX1 3PU}, 
            country={United Kingdom}}

\affiliation[gsi]{
organization={GSI Helmholtzzentrum fur Schwerionenforschung GmbH}, addressline={Planckstrasse 1, 64291}, city={Darmstadt},
country={Germany}
}

\affiliation[liverpool]{
    organization={Department of Physics, University of Liverpool},
    addressline={Brownlow Hill},
    city={Liverpool},
    postcode={L69 7ZX},
    country={United Kingdom}
}

\affiliation[cern]{organization={European Organization for Nuclear Research (CERN)},
            addressline={CH-1211}, 
            city={Geneva 23},
            country={Switzerland}}

\affiliation[lancaster]{
    organization={Department of Physics, University of Lancaster},
    city={Lancaster},
    postcode={LA1 4YB},
    country={United Kingdom}}

\affiliation[ral]{STFC, addressline={Rutherford Appleton Laboratory, Chilton}, city={Didcot}, postcode={OX11 0QX}, country={United Kingdom}}

\affiliation[iceland]{
    organization={Science Institute, University of Iceland},
    addressline={Dunhaga 3},
    city={Reykjavik},
    postcode={IS-107},
    country={Iceland}
}

\affiliation[kth]{
    organization={Division of Space and Plasma Physics, School of Electrical Engineering and Computer Science, KTH Royal Institute of Technology},
    addressline={Teknikringen 31},
    city={Stockholm},
    postcode={SE-10044},
    country={Sweden}
}


\begin{abstract}
A robust and flexible architecture capable of providing real-time analysis on diagnostic data is of crucial importance to physics experiments. In this paper, we present such an online framework, used in June 2025 as part of the HRMT-68 experiment, performed at the HiRadMat facility at CERN, using the Super Proton Synchrotron (SPS) beam line. HRMT-68 was a fixed-target laboratory astrophysics experiment aiming to identify plasma instabilities generated by a relativistic electron-positron beam during traversal of an argon plasma. This framework was essential for experimental data acquisition and analysis, and can be adapted for a broad range of similar-scale experiments with a variety of experimental diagnostics, even those without a standard direct network communication interface. The developed framework's customizable design enabled us to rapidly observe and extract emergent features from a diverse range of diagnostic data. Simultaneously, its modularity allowed for a quick introduction of new diagnostic devices and the modification of our analysis as features of interest were identified. As a result, we were able to effectively diagnose equipment malfunction, and infer the beam's response to varying bunch duration, beam intensity, and the plasma state without resorting to offline analysis, at which time adjustment or improvement would have been impossible. We present the features of this agile framework, whose codebase we have made publicly available so that it may be adapted for future experiments with minimal modification. 
\end{abstract}


\end{frontmatter}

\section{Introduction and Motivation}
\label{Section:Introduction}

Small-scale accelerator experiments often face constraints from limited beam time and resources. As a typical example, the experiments in the CERN's North Experimental area are allocated 1 week of total beam time, which includes setup, installation, beam tuning, data taking, and decommissioning \cite{Banerjee:2774716}. Furthermore, experiments not undertaken at state-of-the-art facilities, for example the European Organisation of Nuclear Research (CERN), will not have access to specialist frameworks, such as CERN's Accelerator Data Logging Service (NXCALS) \citep{sobieszek23:1494}. Smaller experimental facilities which typically accommodate test-beam users and experiments, among them INFN-LNF \cite{Balerna2023} or NCSR Demokritos \cite{NCSRDemokritos_overview}, lack the NXCALS infrastructure, increasing reliance on custom and dedicated equipment to perform online and offline data analysis. Even when such frameworks are present, they may be insufficiently flexible to address experimental needs. In novel platforms like the newly established FIREBALL platform at CERN \cite{arrowsmith21:023103}, experimental demands often change drastically during runtime, and frameworks that are flexible and capable of addressing these changes are essential. 

Though there are conceptually more sophisticated and more reliable frameworks in operation, such as CERN's JCOP/WinCC Open Architecture \cite{GEMDCS_JCOP_Overview}, we have chosen to instead deploy our framework during experimental runtime in order to maintain unlimited control over its functionality, code, and integration. This was essential for continuity between incarnations of the FIREBALL experimental platform, providing us with a framework tailored to our particular experimental requirements, and allowing internal continuity and collaborative modification between subsequent iterations of the experiment. Additionally, our framework was relatively quick to construct, minimally complicated, and highly customizable, and we modified it with ease even during our allocated beamtime to address issues ranging from a camera's optical misalignment to equipment malfunction. Such malfunctions were common and often critical; our Czerny-Turner monochromator \cite{czerny30:792}, tasked with detecting synchrotron emission from within the plasma, ceased operation during experimental runtime. Though the device registered as still operational, it failed to capture images when triggered, and the cause of this failure remains unknown. Additionally, due to damage to an optic fibre, we could not deploy one of our optical components as planned. The code's modular structure, wherein devices can be trivially added or removed from the data extraction protocol, allowed the framework to continue running unimpeded despite these types of failure. On our small collaboration, such obstacles are difficult to resolve due to the lack of technical expertise with respect to specialist equipment.

HRMT-68 differed from its sister experiments at CERN in that it collected a comparatively small amount of data, approximately $20~\text{GB}$ in total. This is in sharp contrast to LHC-centred experiments. For example, during the LHC Run 3, LHCb collected raw data at a rate of $4~\text{TB/s}$ \cite{Fontana2025}. Similarly, ATLAS' Trigger and Data Acquisition (TDAQ) systems will be capable of processing data at an input rate of $4 ~\text{TB/s}$ after the High Luminosity upgrades at the LHC (HL-LHC) \citep{Kopeliansky2023}. Our far smaller scope of data acquisition affected our design considerations, motivating the use of Python, whose performance in our framework is modest and would not be scalable to larger collaborations' experimental needs. 

Particularly in small- or medium-size experiments, the scope of data collection, and the often relatively small fraction of beam spills (or pulses) corresponding to events of interest, motivates the development of software frameworks not only capable of discerning events of interest among a moderate quantity of data, but also of performing analysis and extracting features of interest \textit{during} experimental runtime. This ability to perform data analysis in real-time is of immense value in cases where offline analysis would fail to identify faults such equipment failure or beam misalignment until long after the experiment's conclusion. 

In this technical note, we present such a robust, flexible, and reusable framework for online data analysis of experimental diagnostics, first employed during the HRMT-68 experiment performed at the HiRadMat facility \citep{efthymiopoulos11:1665} at CERN, part of the Super Proton Synchrotron (SPS) accelerator  \citep{arnaldi2025futurefacilitiescernsps}. This was a laboratory astrophysics (see the discussion of recent advances in laboratory astrophysics by Savin {\em et al.}~\cite{savin12:036901}) experiment, which aimed to study plasma instabilities \citep{PhysRevLett.31.1390, Schekochihin2024} produced by an electron-positron pair beam \citep{arrowsmith24:5029} traversing an argon plasma \citep{arrowsmith21:023103}. Our developed framework places an emphasis on readability and modularity, making it easily applicable and reusable for a diverse range of future experiments. Authored in Python, with the option to incorporate a JavaScript Object Notation (JSON) configuration file (see Section \ref{section:code_structure}), it conforms rigidly to the object-oriented programming (OOP) paradigm for maximized encapsulation and abstraction \citep{lutz11b, phillips2015python}, producing a software package that can be easily utilized by those with minimal programming expertise. The resulting framework executes most of its complex data extraction and analysis pipeline without directly interfacing with the user; this ``under-the-hood'' operation is one of the framework's primary strengths. 

The present work is outlined as follows: the motivation and experimental layout of the HRMT-68 experiment is discussed in Section \ref{Section:FireballIII_experiment}, provided to the reader as an exemplar application. We then present our methods of continuous data acquisition in Section \ref{Section:data_collection}, which was crucial in developing a central repository of raw data for online analysis during the experimental run. In Section \ref{section:code_structure}, we discuss the code itself, outlining its hierarchical structure, modular design, and robustness, before quantifying its performance in Section \ref{Section:code_performance}. The paper concludes with a discussion of the framework's strengths and weaknesses, as well as any potential improvements for future incarnations of the code, in Section \ref{section:conclusion}.

\section{The HRMT-68 Experiment}\label{Section:FireballIII_experiment}

The HRMT-68 (FIREBALL-III) experiment utilizes a novel platform recently developed at CERN's HiRadMat facility, which generates an electron-positron pair-beam from the collision of a $440~ \text{GeV}/c$-momentum proton beam with a fixed-target, as detailed by Arrowsmith {\em et al.}~\cite{arrowsmith24:5029,arrowsmith21:023103,Arrowsmith2024thesis}. This produced electron-positron pair-beam passes into a cylindrical argon plasma, produced via inductive discharge \cite{arrowsmith23:P04008, Arrowsmith2024thesis,lieberman25b}. This plasma was contained in a glass plasma cell.

The complexity of this experiment required a diverse collection of diagnostic equipment, which, collectively, had to provide optical, spectroscopic, and magnetic field data from the plasma. A summary of the devices used, and their purpose, is provided in the form of Table \ref{table:devices_summary}. For details on the setup on previous incarnations of the experiment, we point the reader again to refs. ~\cite{arrowsmith24:5029,arrowsmith21:023103,Arrowsmith2024thesis}.

Four screens made of Chromox, a chromium-doped alumina ceramic ($99.5 ~\%$ Al$_2$O$_3$, $0.5~\%$ Cr$_2$O$_3$) \citep{Arrowsmith2024thesis, Burger2016, McCarthy2003}, were used in the setup. Chromox produces scintillation light in the visible spectrum at wavelengths of $691~ \text{nm}$ and $694~ \text{nm}$ \citep{Arrowsmith2024thesis}. Each of the four screens was imaged individually using four cameras. These four cameras, designated $\text{HRM}\{3, 4, 5, 6\}$, were collectively dubbed the ``Chromox cameras''. Cameras $\text{HRM}\{3, 4\}$ imaged the plasma cell directly, and were referred to as the ``plasma cell'' cameras; the two other Chromox cameras imaged screens on the far side of a dipole magnet, and were dubbed the ``spectrometer cameras''. Deflection of particles in a dipole field is a function of their energy, and hence scintillation patterns produced by deflected electrons/positrons as they impinge these two latter screens provide a measure of the electron/positron energy spectrum. Two screens were necessary, as electrons and positrons are deflected in opposite directions by a dipole magnetic field.

For any initial velocity, charged particles emit optical transition radiation (OTR) when passing between two media of differing dielectric constant; the variation in electric and magnetic fields across this boundary results in this radiative emission \cite{otr_dasilva, otr_gitter}. The collection of photons produced via OTR is often used in beam diagnostics to characterize the longitudinal density profile of accelerator beams \citep{Chen2023, Fedorov_2020}, as was done in the HRMT-68 run. OTR was produced by introducing a copper foil into the path of the beam, with the subsequent emitted radiation collected by a ``streak unit'' \citep{hamamatsu} with an adjustable slit width. The information can be encoded in a 2D image, which contains information about the pulse duration of incident radiation, and is captured by a digital camera; such a digital camera was mounted on the streak unit during the experiment. 

Synchrotron radiation is produced by the deflection of charged particles in magnetic fields \citep{Walker1994}. In the HRMT-68 experiment, synchrotron emission was used to infer the interaction of the electron-positron beam with the magnetic fields produced in the plasma. Synchrotron radiation emission from the plasma was collected using a charge coupled device (CCD) \citep{Hui2020} camera (Andor iStar \citep{andor_istar}) mounted on a Czerny-Turner monochromator \citep{czerny30:792}.

Magnetic field diagnostics for the plasma were provided by three B-dot probes, named for the overdot denoting differentiation of the magnetic field with respect to time: $\dot{\mathbf{B}}$ \citep{Bose_2018}. These probes measure the magnetic field strength within the plasma via a practical application of Faraday's Law: electromotive force (e.m.f.)\ induced in the coils of the B-dot probe is proportional to the rate of change of magnetic flux through the coils, using Equation~\ref{eq:bdot_eq} (see refs.~\cite{Griffiths_2017}). \begin{equation}
    \varepsilon = -\frac{\text{d} \phi}{\text{d} t} = - N \frac{\text{d}}{\text{d}t} \int_S \vec{dS} \cdot \vec{B}  = -N \int_S \vec{dS} \cdot \dot{\vec{B}}
    \label{eq:bdot_eq}
\end{equation}
The e.m.f. induced in the B-dot coils was measured using two identical four-channel oscilloscopes (Tektronix 6 Series MSO \citep{textronix6}).

\begin{table}
\centering
\begin{adjustbox}{max width=1.01\textwidth}
\begin{tabular}{|ll|l|}
\hline
\multicolumn{2}{|c|}{\textbf{Device}}                                         & \multicolumn{1}{c|}{\textbf{Purpose}}                                                                                                                                                   \\ \hline
\multicolumn{1}{|l|}{\multirow{2}{*}{Chromox Cameras}} & Plasma Cell Cameras  & Observe transverse profile of electron-positron pair-beam.                                                                                                                \\ \cline{2-3} 
\multicolumn{1}{|l|}{}                                 & Spectrometer Cameras & Observe energy spectrum of electrons \& positrons exiting plasma cell.                                                                                                                   \\ \hline
\multicolumn{2}{|l|}{OTR Streak Unit + Camera}                                & \begin{tabular}[c]{@{}l@{}}Produce spatially- and temporally-resolved information about\\ OTR produced by pair-beam.\end{tabular}             \\ \hline
\multicolumn{2}{|l|}{Synchrotron Emission Spectrometer + Camera}              & \begin{tabular}[c]{@{}l@{}}Resolve wavelengths of radiation present in synchrotron emission\\ due to electron-positron beam’s deflection by plasma $\textbf{B}$-field.\end{tabular} \\ \hline
\multicolumn{2}{|l|}{B-dot Probes + Oscilloscopes}                            & Observe/plot $\textbf{B}$-field induced in plasma.                                                                                                                                      \\ \hline
\end{tabular}
\end{adjustbox}
\caption{Summary table of devices present in the HRMT-68 experiment. Care is made to distinguish the spectrometer and streak unit from the two cameras (one per unit) mounted on them; raw data collected from the experiment was from these cameras, rather than the spectrometer and streak unit themselves.}
\label{table:devices_summary}
\end{table}

\section{Data Collection}\label{Section:data_collection}

Collection and logging of device data was potentially the most challenging aspect of the online framework. Data from each of the experiment's diagnostics (enumerated in Section \ref{Section:FireballIII_experiment}) was collected into a centralized repository, with the online analysis framework constructed to accommodate this and read each device's data from a specified location (see Section \ref{section:code_structure}). 

\subsection{Construction of a Centralized Data Repository}

CERN's sophisticated Accelerator Data Logging Service (NXCALS) \citep{sobieszek23:1494} necessitates the existence of a special front-end, designated ``Front-End Software
Architecture'' (FESA)\footnote{See \href{https://confluence.cern.ch/spaces/viewspace.action?key=FESA3}{\url{https://confluence.cern.ch/spaces/viewspace.action?key=FESA3}}.}. FESA was developed for LHC and its injectors, and is extremely versatile and adaptable. However, its implementation requires specialist expertise usually not present on small-scale international collaborations, and many experiment-specific pieces of equipment are not trivially compatible with FESA. For example, the HRMT-68 used a NAVIO matching network \cite{AdvancedEnergy2018Navio} for impedance matching when delivering RF power to the plasma cell; integrating the matching network into FESA would have required time and expertise that was unavailable to the collaboration. 

Finally, it is generally known that the development of these generalized software frameworks (e.g.\ NXCALS/FESA) is driven by the global needs of each organization and its flagship experiments, leaving little to no space (in terms of resources or support) for the requirements and the versatility necessary for smaller experiments to be taken into account. This was our primary motivation for this work.

The variety of data from a range of diagnostics (Table \ref{table:devices_summary}) motivated the construction of an easily-accessible, experiment-specific central repository of data collected during the HRMT-68 campaign. Similar small-scale experiments may only have access to a generic cloud-based storage service accessible over a network connection, as was the case on HRMT-68. Therefore, our framework was designed to cater to this limitation.

In our case, we profited from CERN's online Jupyter environment, Service for Web-based ANalysis (SWAN, described in the CERN SWAN documentation ~\cite{swan_docs}), which allows users to run code from any location. The Jupyter environment provides a user-friendly front-end for executing Python without the need for SSH connections to remote servers or installation of local versions of Python. SWAN is directly integrated with CERN's EOS Open Storage (EOS)\footnote{EOS acronym is recursive.}/CERNbox online file storage configuration; EOS is CERN's disk storage system, with a storage capacity of $\sim780~\text{PB}$ \citep{EOS_docs}, and ``CERNbox'' is CERN's cloud-based file-sharing and storage service \citep{CERNbox_docs}. Therefore, the use of SWAN permitted any user with EOS/CERNbox read and write permissions and an internet connection to run our software from any location. 

Data from each diagnostic was assigned a unique directory in this central data repository, which was pointed to by the code's configuration (as described in Section~\ref{section:code_structure}). However, this framework would work equally well with any cloud-based or local file storage service.

In order to synchronize the different device data to the repository, each device was configured to either save data directly into CERNbox via a mounted network drive, or to a local directory synced continuously with the appropriate device directory on CERNbox, depending on the device. The speed of the synchronization process varied greatly between devices, from $\lesssim 5$ seconds for devices with access to a mounted network drive, to $O(10)$ seconds for devices whose local directories were synced remotely.

\subsubsection{Challenges with Timestamp Tagging}\label{section:timestamp_challenges}

A central goal of the framework was the creation of a \textit{consistent} scheme for logging all data acquired during the experiment. During the HRMT-68 experiment, several experimental conditions were continually modified. These included power supplied to the plasma cell, position of the graphite target used to produce the electron-positron beam, as well as the beam density requested from the SPS. Keeping track of which data files corresponded to which experimental conditions was therefore essential in subsequent data analysis. Therefore, we aimed to maintain a centralized ``shot log'' which tracked all experimental conditions during each beam extraction, and the time of each extraction. The aim was to then compare these extraction times to the time(s) at which each device was triggered. 

\begin{figure}[ht]
    \centering
    \includegraphics[width=\linewidth]{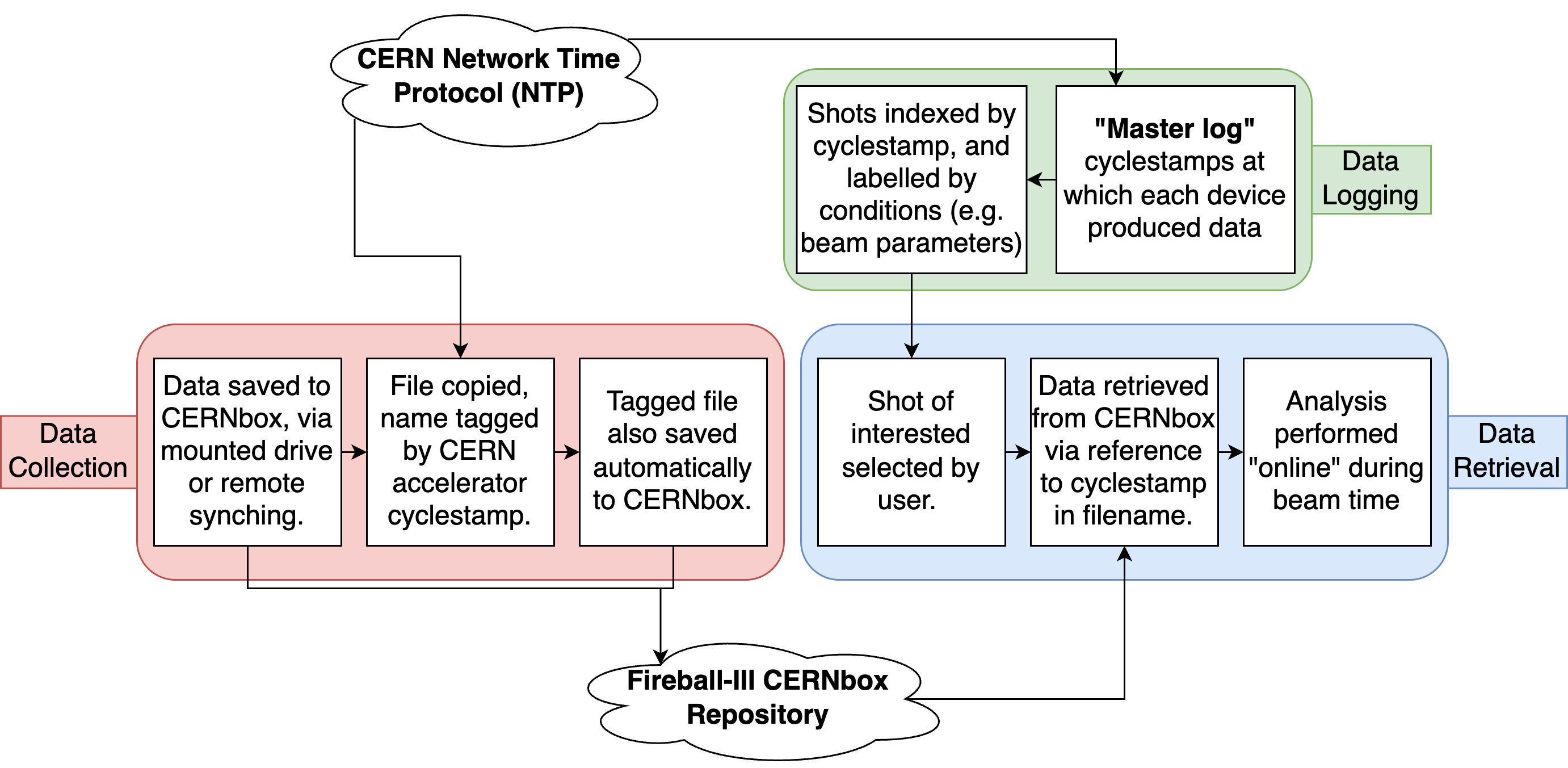}
    \caption{Schematic diagram showing the \textbf{ideal} (and initially-proposed) data tagging, logging, and retrieval processes. In reality, the shot log could not be automatically generated, and the process of time-stamping data using the CERN accelerator cyclestamp proved infeasible due to instabilities in network connection.}
    \label{fig:ideal_collection}
\end{figure}

Initially, the process of sorting devices' data via timestamp was to be carried out by automatically generating duplicates of the devices' output data files and appending a UNIX timestamp \citep{louis2020time} to the filename. This proposed process is outlined in Figure \ref{fig:ideal_collection}. The timestamp used to label device data was to be provided by CERN's Network Time Protocol (NTP) \citep{K2023}, and corresponds to the timestamp used to index CERN's accelerator cycles. This timestamp is termed the ``cyclestamp''. During our data collection and logging procedures, when relying on a connection the CERN NTP server, we observed up to an $O(10~\text{s})$ discrepancy between the timestamps associated with each devices' data for the same extraction. This limitation was due to delays in the devices' data being saved to our cloud-based repository. However, this observed precision was more than sufficient to remove ambiguity surrounding which data corresponded to which extraction, as intervals between subsequent extractions from the beamline during the experimental run were around 90 seconds.

However, this initial method of tagging the filename with a timestamp directly proved infeasible for several reasons: firstly, it necessitates duplication of all data, resulting in a doubling of storage requirements. Given that hard disk space is often at a premium in small-scale experiments, this complication is unsatisfactory and will often be unacceptable. Secondly, timestamping the files (by creating a copies of each file with a modified filename) relied on continuous execution of a Python script which maintains a connection to the CERN NTP server via a WebSocket connection \cite{rfc6455}, which increased the framework's vulnerability to endemic network latency and connectivity concerns. As a result, only a fraction of collected data was correctly timestamped. We retain a description of this idealized process in order to motivate future efforts to dynamically tag data in such an unambiguous fashion while addressing storage and connectivity concerns.

\begin{figure}[ht]
    \centering
    \includegraphics[width=0.85\linewidth]{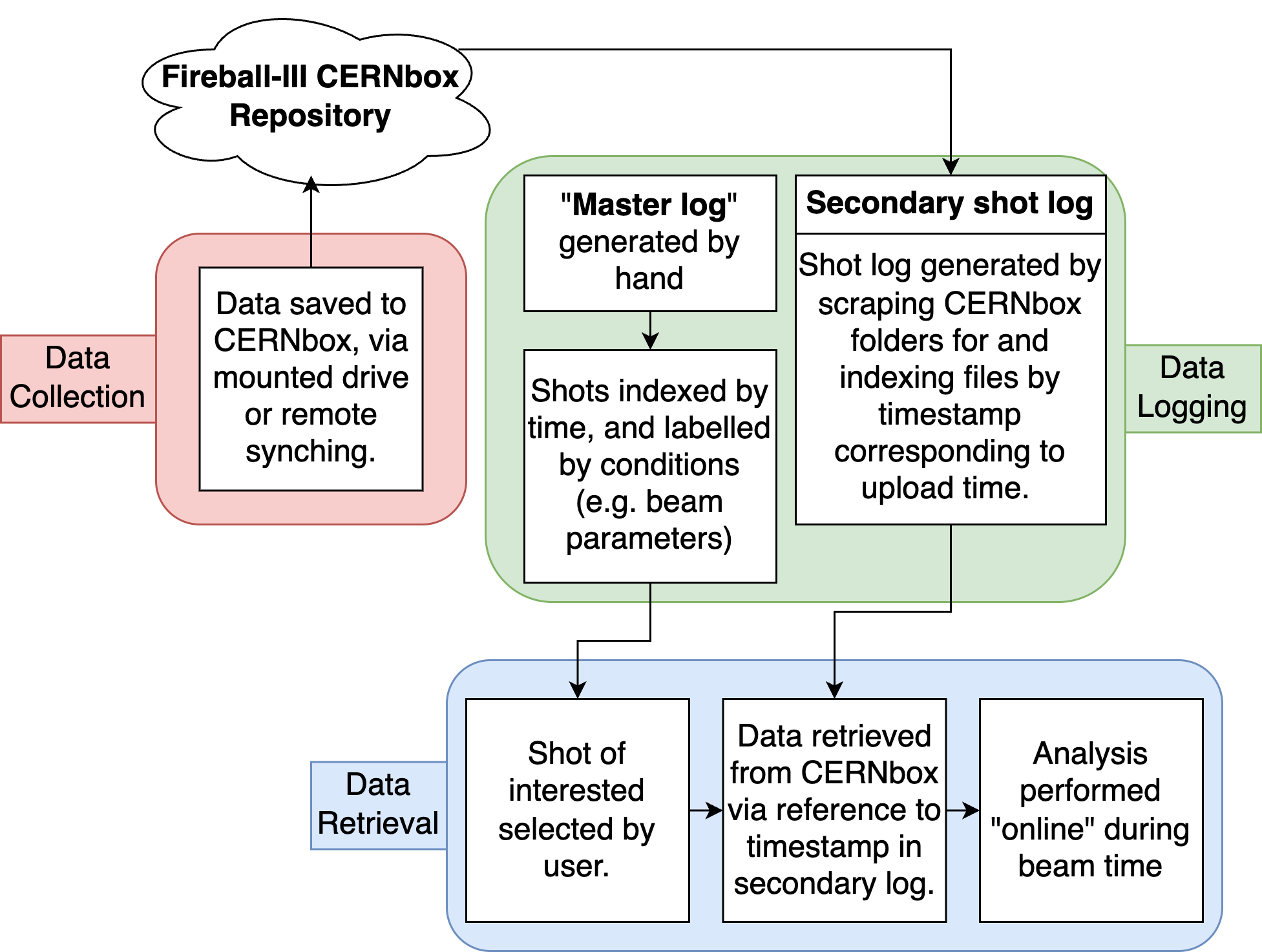}
    \caption{Schematic diagram showing the system of data tagging, logging, and retrieval \textbf{in reality}, without tagging via cyclestamp. Here, a secondary log with minimal human-readable detail is introduced.}
    \label{fig:reality_collection}
\end{figure}

A revised scheme (shown in Figure \ref{fig:reality_collection}) was introduced to compensate for the difficulties that arose while attempting to implement the scheme shown in Figure \ref{fig:ideal_collection}. In place of an automated log of timestamps at which each device triggered, a simplified ``secondary shot log'' was maintained, which indexed the data from each device by the UNIX timestamp corresponding to the upload time on CERNbox rather than by relying on the cyclestamp. Unlike the central ``master'' log, which had to be maintained by hand and contained details of experimental conditions, the secondary log contained no detail of experimental conditions, such as target placement or conditions of the plasma. However, the secondary shot log was automatically generated, and hence resilient to human error. 

This process, while cumbersome, was internally consistent and resulted in the creation of two centralized shot-logs for all diagnostic devices, where data files were indexed by a universal timestamp key. This eliminated ambiguity as to which data files corresponded to which experimental conditions and which beam extraction. These timestamp keys can then be fed into the framework's frontend to filter for data of interest across all devices.

Our framework retains the ability to access data for either method, both the case in which the filename is explicitly modified and appended with the UNIX timestamp, or the case in which data is logged by the time at which it first appeared on the EOS/CERNbox file system. The former case is preferable, as it provides all data with a permanent timestamp which is resilient to file modification or migration of data off the EOS/CERNbox storage system; both cases will result in changes to the files' modification time metadata, which was used in our revised method (Figure \ref{fig:reality_collection}) to index data.

\section{Code Structure and Design}\label{section:code_structure}

Having resolved the issue of indexing data by the time of acquisition, allowing us to filter out shots of interest, the subsequent goal of this project was to construct a set of analysis methods to be performed on the output data across all devices. Our desire to do so ``online'' meant that these methods had to be executable as soon as the data was available on our centralized CERNBOX/EOS repository. This demanded both flexibility (the ability to introduce and remove diagnostics at will) and specificity (with each device requiring its own unique analysis and processing). The codebase was written in Python 3.11.9. 

The choice of Python resulted in somewhat of a compromise on speed and efficiency. However, it provided equally significant gains in flexibility. Python is by some metrics the most widely-used programming language \cite{TIOBE_Index}; hence our framework necessitates no specialist skills other than basic proficiency in a common programming language, compared to the specialist expertise required for device integration into NXCALS/FESA. This guarantees a baseline level of longevity (Python is unlikely to exit use in academic circles in the near future) and accessibility. Moreover, Python offers an extensive array of packages specialized in data extraction and analysis, such as the popular Pandas package \cite{reback2020pandas} and the Numpy library, whose performance is optimized via the use of an underlying C/C++ backend for accelerated computation \cite{harris2020array}. 

\begin{figure}
    \centering
    \includegraphics[width=1\linewidth]{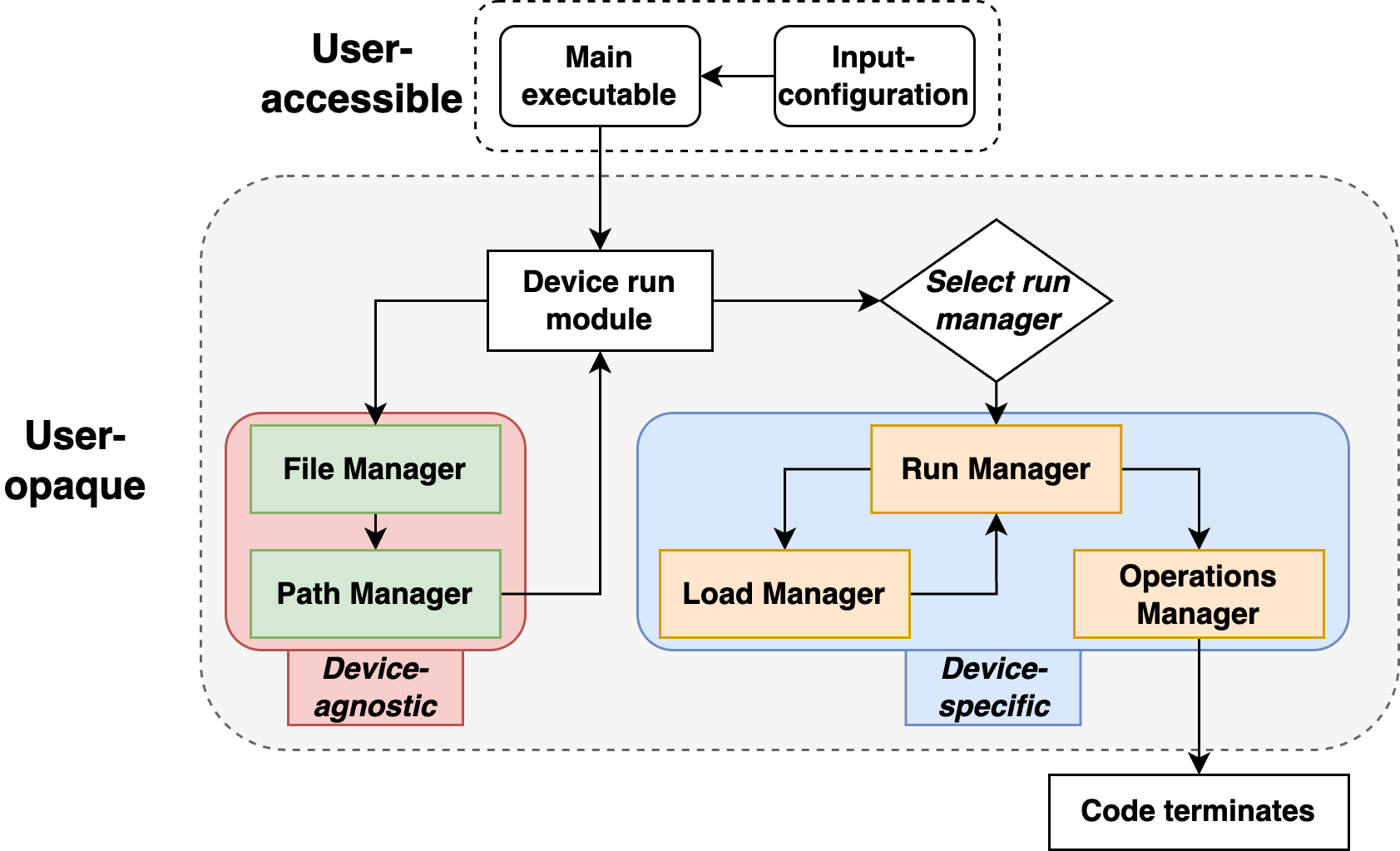}
    \caption{Schematic tree of the framework's hierarchical structure, with modules shown in square boxes. Care is taken to distinguish device-agnostic modules (responsible for collecting file names from directories) from device-specific ones, where derivative classes for different device species were implemented to account for differing file formats and extensions, number of channels of data, and different plotting/analysis requirements for each device.}
    \label{fig:code_tree}
\end{figure}

The code's hierarchical structure, shown schematically in Figure \ref{fig:code_tree}, was chosen to allow a high degree of modularity, addressing both the issue of flexibility and the need to specify analysis individually for each device. Its high degree of modularity and class-based structure also allowed members of the collaboration not directly involved in the code's development to sample code piecewise from the analysis pipeline to perform their own specialized analysis in parallel with the online data collection.

\subsection{Configurability}\label{Section:Code_Configurability}

\begin{figure}
    \centering
    \includegraphics[width=1\linewidth]{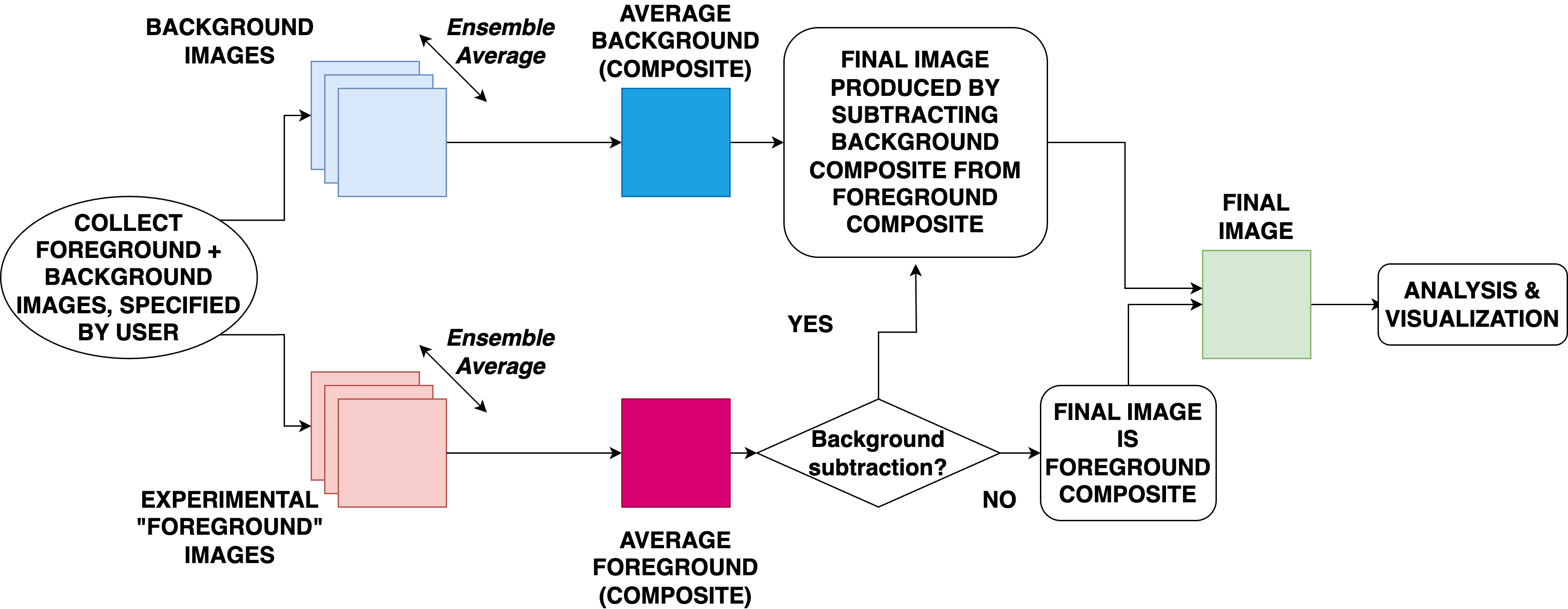}
    \caption{Schematic showing the code's image averaging pipeline, including how multiple background shots could be combined into a single ``composite background''. Multiple foreground shots could also be combined into a ``composite foreground'' shot. Subsequent analysis is performed on the output of the subtraction between the composite images.}
    \label{fig:background_subtraction}
\end{figure}

The code was designed to read commands from an input configuration (either via JSON file or a manually-edited ``input configuration'' dictionary in the Python environment) to maximize readability and customization. The JSON format is simple, has minimal syntax, and closely resembles plain English. The purpose of the configuration file was to reduce the proliferation of hard-coded values in the framework which would need to be modified between runs, a process prone to human error. In the input configuration, users could specify shot numbers of interest and/or request statistical averaging across these shots. As depicted in Figure \ref{fig:background_subtraction}, the user is able to specify ``background'' data to be subtracted from data of interest. The configuration allowed for displaying images without performing any analysis to expedite execution speed; quantitative assessment of how this contributed to the code's execution speed can be found in Table \ref{table:code_speed}.

\subsection{Modularity}
The code's modular structure, depicted in Figure \ref{fig:code_tree}, focusing on the image analysis functionality, was key to its adaptability. As experimental goals were modified, or regions of interest identified, quick refactors could be made to analysis methods without disrupting the data collection and extraction processes. This allowed for a ``plug-in-and-play'' model: collaborators external to the project could implement their own analysis code, done independently and without any knowledge of the HRMT-68 online analysis framework, by simply inserting their extraction and analysis methods into the relevant modules in the framework. A detailed description of the different key-modules are described below.

The \textbf{file manager} module is tasked with creating an index of file \textit{names} and their corresponding upload time, and this index is converted into a time-ordered index of file \textit{paths} by the \textbf{path manager}, as described in Section \ref{Section:data_collection}. The \textbf{run} and \textbf{load manager} modules have two subclasses, one each for handling time-domain waveform data from oscilloscopes, and for handling bitmap image data from cameras. All device-specific analysis during runtime is controlled by the \textbf{run manager}, including statistical averaging and error propagation. The \textbf{load manager} subclass that the code calls is device-specific to account for the different file types and data logging structure of each device. All analysis and data visualization after the relevant data is extracted and preprocessed is handled by the \textbf{operations manager}, after which the code terminates. 

This hierarchical substructure is ideal as it allows for the straightforward, ``plug-in'' implementation of analysis methods for additional diagnostic devices, as required. As new devices are introduced to an experimental setup, desired analysis methods, such as visualization, cropping, and marginalization over an axis, can be implemented in the \textbf{operations manager} module. This can be done without modification to  the \textbf{path}, \textbf{run}, and \textbf{load} manager classes, which contain generic functionality for logging and extracting data from any device.

\section{Performance and Results}\label{Section:code_performance}

In this section, we quantify the code's performance when producing analysis and/or visualization of diagnostic device data. For these tests, we tested the code with data from: the OTR streak unit camera, synchrotron spectrometer camera, a single Chromox camera (a single plasma cell camera, see Section \ref{Section:FireballIII_experiment}), and an oscilloscope reading output from one of our B-Dot probes (see Section \ref{table:devices_summary}). 

In Section \ref{subsection:performance_results}, we present a summary of the code's performance in Tables \ref{table:code_speed} and \ref{table:peak_mem}, where we measure execution time and peak memory usage, respectively. The tests were performed over a remote connection to a CPU (Intel(R) Xeon(R) Silver 4216 CPU, 2.10 GHz \citep{intel_xeon}) on the CERN cluster, running an AlmaLinux9 OS distribution \citep{almalinux}, via CERN's SWAN service, with a cumulative 8 GB of RAM and 2 cores allocated to the session. The ethernet connection of the remote CPU was configured to a ceiling of 10000 Mbps, and measurement of computer's download speed yielded $\sim 200 ~ \text{MB/s}$.  

All tests were performed with the background subtraction routine to maximize computational complexity. This was produce estimates for a ``worst-case'' run time during testing. Background subtraction was done by averaging over 5 shots designated as ``background'' to create a ``composite background'' image, which was then subtracted from the foreground image (see Figure \ref{fig:background_subtraction}). Summaries of pixel dimensions for the three types of device producing visual data are provided in Table \ref{tab:pixel_dims} 

\begin{table}[]
    \centering
\begin{tabular}{@{}lccc@{}}
\toprule
\multicolumn{1}{c}{\textbf{Device}} & \textbf{\begin{tabular}[c]{@{}c@{}}Image \\ Pixel Dimension\end{tabular}} & \textbf{\begin{tabular}[c]{@{}c@{}}Approx. Size \\Per Image \\ / MB\end{tabular}} & \textbf{\begin{tabular}[c]{@{}c@{}}No. \\ of Shots \\ Collected\end{tabular}} \\ \midrule
Synchrotron Spectrometer            & $1024 \times 1026$                                                        & $4.2$                                                                 & $\approx 100$                                                                 \\
OTR Streak Camera                   & $2049 \times 2049$                                                        & $14.5$                                                                & $\approx 300$                                                                 \\
Chromox Camera                      & $349 \times 605$                                                          & $5.0$                                                                 & $\approx 1000$                                                                \\ \bottomrule
\end{tabular}
    \caption{Dimensions of images, in pixel number, produced by each of the three devices whose data was used to test the framework's performance.}
    \label{tab:pixel_dims}
\end{table}

Each entry of execution time in Table \ref{table:code_speed} is an average over 10 trials. 

\begin{figure}
    \centering
    \includegraphics[width=\linewidth]{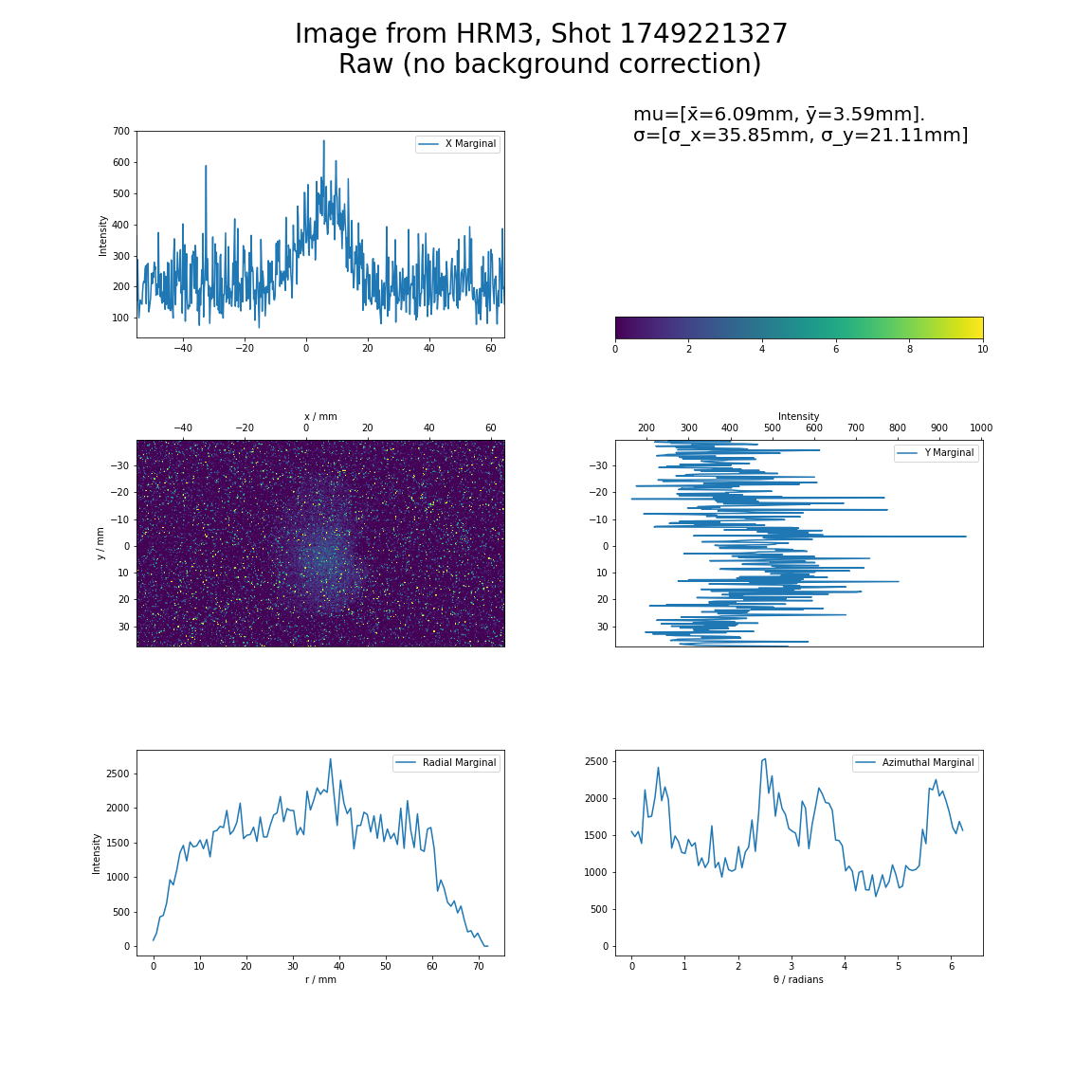}
    \caption{Sample output from one of the Chromox cameras, HRM3 (see Section \ref{Section:FireballIII_experiment}), tasked with monitoring the upstream cross-section of the plasma cell.}
    \label{fig:chromox_sample}
\end{figure}

As mentioned previously in section \ref{Section:Code_Configurability}, the code could have its configuration modified between runs. Users could specify that the framework perform no analysis and simply plot device data in order to increase execution speed in cases where it was only necessarily to visualize the camera images or oscilloscope traces, and we include such test cases in our performance evaluation. We provide a sample output from the analysis framework for one of the Chromox cameras (see Section \ref{Section:FireballIII_experiment}), HRM3, in Figure \ref{fig:chromox_sample}.

Results for tests of peak memory usage in Table \ref{table:peak_mem} were averaged across 10 runs of the code. This was measured using Python 3's inbuilt memory profiler/memory usage function. 

In order to account for the significant reduction in the code's speed introduced by calling analysis methods on the Chromox cameras, additional tests to quantify two of the most computationally-expensive analysis methods involved in the Chromox calculations were performed. These computationally expensive methods included calculating the first and second moments of image pixel intensity (i.e.\ the centroid of pixel intensity, and standard deviations on this centroid in both dimensions), as well as coordinate transformations from $(x, y)$ 2D Cartesian coordinates to $(r, \theta)$ plane polar coordinates. The latter method was performed in order to generate radial and azimuthal marginals. These are summarized in Table \ref{table:chromox_speed}.

\subsection{Results and Discussion}\label{subsection:performance_results}

\begin{table}[ht]
\centering
\begin{adjustbox}{max width=\textwidth}
\begin{tabular}{|l|c|c|c|c|}
\hline
\textbf{Device} & \multicolumn{4}{c|}{\textbf{Code Execution / s}} \\
\cline{2-5}
& \shortstack{\\\textbf{Single shot}\\\textbf{No Analysis}} 
& \shortstack{\\\textbf{Single shot}\\\textbf{With Analysis}} 
& \shortstack{\\\textbf{Average shot}\\\textbf{No Analysis}} 
& \shortstack{\\\textbf{Average shot}\\\textbf{With Analysis}} \\
\hline
Chromox Camera & 
 $1.7 \pm 0.1$& 
 $10.8 \pm 0.3$& 
 $2.1 \pm 0.2$ & 
 $12.1 \pm 0.4$\\
\hline
Synchrotron Spectrometer Camera& 
 $6.7 \pm 0.3$& 
 $6.3 \pm 0.6$& 
 $9.5 \pm 0.6$& 
 $10.4 \pm 0.5$\\
\hline
OTR Streak Unit Camera & 
$24.8 \pm 0.6$& 
 $22.8 \pm 0.4$& 
$37.9 \pm 0.7$& 
 $41 \pm 2$\\
\hline
B-dot Probe Oscilloscope & 
$3.6 \pm 0.9$& 
$3.9 \pm 0.5$& 
$3.5 \pm 0.5$& 
$5.0 \pm 0.8$\\
\hline
\end{tabular}
\end{adjustbox}
\caption{Time-of-execution for the code for each species of device. Tests were done for cases with (``average shot'') and without (``single shot'') foreground averaging (see Figure \ref{fig:background_subtraction}). For either case, trials were performed both with and without calling analysis on the resulting background-corrected data. ``No-analysis'' cases simply visualize the devices' data. Background subtraction was performed in all four cases, with a ``composite background'' image/oscilloscope trace produced by averaging five background images/traces. Values are statistical averages over 10 runs of the code.}
\label{table:code_speed}
\end{table}

\begin{table}[ht]
\centering
\begin{adjustbox}{max width=\textwidth}
\begin{tabular}{|l|c|c|c|c|}
\hline
\textbf{Device} & \multicolumn{4}{c|}{\textbf{Average Peak Memory Usage / MB}} \\
\cline{2-5}
& \shortstack{\\\textbf{Single shot}\\\textbf{No Analysis}} 
& \shortstack{\\\textbf{Single shot}\\\textbf{With Analysis}} 
& \shortstack{\\\textbf{Average shot}\\\textbf{No Analysis}} 
& \shortstack{\\\textbf{Average shot}\\\textbf{With Analysis}} \\
\hline
Chromox Camera & $536.5 \pm 0.1$ & $463.1 \pm 0.3$ & $587.1 \pm 0.2$ & $618.8 \pm 0.1$ \\
\hline
Synchrotron Spectrometer Camera & $537 \pm 1$ & $470 \pm 10$ & $589 \pm 2$ & $630 \pm 30$ \\
\hline
OTR Streak Unit Camera & $918 \pm 4$ & $890 \pm 20$ & $976 \pm 1$ & $1030 \pm 20$ \\
\hline
B-dot Probe Oscilloscope & $536.6 \pm 0.2$ & $463.0 \pm 0.2$ & $587.8 \pm 0.1$ & $619.0 \pm 0.2$ \\
\hline
\end{tabular}
\end{adjustbox}
\caption{Average peak memory usage in megabytes (MB) for each device in the four test scenarios, identical to those explored in Table \ref{table:code_speed}. Values of peak memory usage are statistical averages over observed peak usage across 10 runs of the code.}
\label{table:peak_mem}
\end{table}

\begin{table}
\centering
\begin{adjustbox}{max width = \textwidth}
\begin{tabular}{|c|c|c|}
\hline
\textbf{\begin{tabular}[c]{@{}c@{}}Average Total Execution Time \\ for Chromox Run / s\end{tabular}} & \textbf{\begin{tabular}[c]{@{}c@{}}Average Total Execution Time,\\ Moment Calculation Methods / s\end{tabular}} & \textbf{\begin{tabular}[c]{@{}c@{}}Average Total Execution Time,\\ Coordinate Transformation Methods / s\end{tabular}} \\ \hline
$10.6 \pm 0.3$                                                                                           & $2.8 \pm 0.1$                                                                                                   & $3.39 \pm 0.05$                                                                                                           \\ \hline
\end{tabular}
\end{adjustbox}
\caption{Execution times for a full run of the analysis code on Chromox images from the tested Chromox camera, as well as the Chromox camera moment and coordinate transformation methods. These results are averages over 10 timed trials of code execution for ``single shot'' mode on one of the Chromox cameras.}
\label{table:chromox_speed}
\end{table}

From the tests of code speed displayed in Table \ref{table:code_speed}, we conclude that the foreground-averaging in the absence of analysis methods introduces negligible additional overhead for the Chromox camera and B-dot probe data. In the former case, execution time for foreground-averaging only introduces a $\sim (24\pm12) ~\% $ increase in execution time, whereas effects to execution time for displays on the B-dot probe oscilloscope execution time are statistically negligible, and indeed execution time actually decreased, changing by $\sim (-3\pm30)~\%$ when averaging was introduced. We conclude that the code's statistical averaging methods are not the limiting factor, and that instead variations in connection/download speed between clusters accessed via SWAN and the CERNbox service, affected by network load and user demand, are far more likely to be responsible. 

Inclusion of analysis methods, however, did introduce significant overhead in the execution process; in the case of the Chromox camera images: this resulted in percentage increases of  $\sim (540 \pm 40)~\%$ and $\sim (480\pm60)~\%$ in execution time for the ``single shot'' and ``average shot'' cases, respectively. The methods used to calculate first- and second-order moments are computationally-costly, iterating over all pixels in the image and therefore scaling as $N^2$ for an image of pixel dimension $N$. Transforming the images' pixels from Cartesian to planar polar coordinates also shared this computational complexity. As shown in Table \ref{table:chromox_speed}, the execution times for these two functions were investigated individually. Cumulatively, they accounted for $(58 \pm 2) \%$ of the entire runtime for calling analysis on the Chromox images.

For the B-dot probe oscilloscope analysis, percentage increases for execution time with the introduction of analysis was $\sim (8 \pm 30)~\%$ in the ``single shot'' case, and $\sim (43\pm30)~\%$ in the ``average shot'' case. The large confidence intervals on these results is a reflection that run-to-run fluctuations in execution speed were a significant fraction of the mean execution speed.

Operations with the streak unit camera were clear outliers in both execution time (Table \ref{table:code_speed}) and memory consumption (Table \ref{table:peak_mem}). This is likely due to the larger pixel dimensions of the images from the synchrotron spectrometer and OTR streak unit cameras ($1024\times1026$ and $2049 \times 2049$ pixels per image, respectively) compared to the dimensions of the single Chromox camera which was used during code testing ($349 \times 605$ pixels). 

Despite the increase in execution speed introduced by the larger pixel count in the spectrometer and streak unit images, introducing analysis resulted in only marginal increases in execution time. For the spectrometer camera, introducing analysis resulted in minor variation in execution time ($(-6 \pm 10)~\%$ (i.e.\ a \textit{decrease}) for the ``single shot'' case, and $(9 \pm 9) ~ \%$ increase for the ``average shot'' case). 

For streak unit analysis, variation between cases with and without analysis was larger ($(-8 \pm 3) ~ \%$ variation for the ``single shot'' case (once again, a \textit{decrease} in execution time), and $(8 \pm 6) ~\%$ increase for the ``average shot'' case). It is worth noting that the comparatively large variation between cases with and without analysis for the ``single shot'' case is unlikely representative of inefficiencies in the code; execution time was shown to actually decrease when analysis was introduced, despite the fact that operations performed by the code without calling analysis are a subset of operations performed when analysis is requested. This is most likely due to variations in network conditions.

We would also like to highlight that, for all devices, the analysis methods incur only a small amount of memory cost; indeed, memory consumption during tests across all devices actually decreased between the ``single shot'' cases when analysis was introduced; owing to the fact that there was an increase in memory consumption after the addition of analysis for the ``average shot'' cases, and the fact that the drop in peak memory consumption occurred for all devices, we suspect that this may be due to changes in the Jupyter kernel state between tests, though the exact mechanism explaining this occurrence remains unknown. This may also explain the decrease in execution time when analysis was introduced in the ``single shot'' cases for the OTR streak unit camera and spectrometer cameras, which are difficult to analyze due to the increased number of operations performed in the analysis case.

\subsection{Evaluation}

Though our framework's performance is not state-of-the art, its ability to address limitations on data collection (as described in Section \ref{Section:data_collection}) and cater to a wide range of diagnostics (Section \ref{Section:FireballIII_experiment}) while maintaining a high level of flexibility and customizability made it invaluable to the experimental runtime. Our decision to implement our analysis methods directly in the framework's codebase also served as an invaluable, internal ``alpha test'' phase of our analysis scripts. For example, we were able to identify regions and shots of interest in preparation for our extensive offline data analysis efforts; thus, such specialist frameworks have significant potential in providing small-scale experimental collaborations with a ``running start'' on their analysis for subsequent publication.

\section{Conclusions and Future Work}\label{section:conclusion}

In this work, we have presented an online framework which is flexible to varying demands from a user, and which is moderately fast even under comparatively adversarial circumstances. It is highly modular, allowing easy implementation of data extraction and analysis code for additional diagnostic devices. Though the code's performance during experimental runtime was largely successful, we would like to also highlight the framework's weaknesses, and how it could be improved and modified in the future.

\subsection{Introducing Caching}
Improvements could be made to execution speed by locally caching data from shots of interest. Often, it was desirable to plot data from the same shot multiple times in order to, for example, crop to regions of interest or adjust image contrast for visibility. This required multiple code runs, with each run limited by the speed at which data could be acquired over a network connection. In future, it would be useful to introduce an optional ``caching'' feature, allowing users to store data from a desired shot locally over a short period, with the option of ``clearing'' the cache when this data was no longer required.

\subsection{Tagging by Cyclestamp}
Experimental collaborations using the framework should aim to have some universally-recognized system of logging data by acquisition time. At CERN, as discussed in Section \ref{Section:data_collection}, a strong candidate would be the UNIX timestamp corresponding to each accelerator cycle (``cyclestamp''), as data filenames can be ``stamped'' by unique UNIX timestamp corresponding to beam extraction times. The framework codebase is easily adjusted to allow indexing by UNIX timestamp so long as it knows the position of the timestamp in the filename string. More sophisticated standards of synchronization are possible and preferable to the kind employed in this framework. For example, the White Rabbit project at CERN provides sub-nanosecond synchronization precision using the IEEE 1588-2008 Precision Time Protocol (PTP) \cite{whiterabbit6070148}. Future experiments may wish to distinguish between multiple data acquisition events for the same beamline extraction if devices are expected to trigger more than once at the sub-second level.

However, in the construction and execution of our framework, bottlenecks in speed and performance emerged primarily due to delays in the process of saving files from the multitude of diagnostic devices (described in Section \ref{Section:FireballIII_experiment}) to our cloud-based CERNBox/EOS storage system, and this delay was folded into our latency budget constraints. Our framework has no method for increasing the speed of caching diagnostic data to the CERNBox/EOS repository after beam extraction, which is a hardware performance concern rather than synchronization rate limitation. Modifying our framework to include sub-nanosecond synchronization schemes will fail to address this shortcoming. Indeed, during our experiment, timestamp resolution needed only to be lower than the $90~\text{s}$ interval between beamline extractions (see Section \ref{section:timestamp_challenges}). Synchronization methods with precision on or below the order of $1~\text{s}$ are readily achievable using standard operating system methods for synchronization between devices connected over Wi-Fi, such as the Windows Time (W32Time) Service \cite{MS_SupportBoundaryHighAccuracy}, which employs NTP and can achieve $O(1~\text{s})$ synchronization if network latency does not exceed $100~\text{ms}$ between devices. 

Further improvements to the framework should focus on increasing the understanding of the underlying hardware, particularly the precise rate at which each diagnostic device logs data to the CERNBox/EOS repository, and the latency between beam extraction and data caching which subsequently emerges for each device. Revising the framework to give increased understanding and control over the underlying diagnostic hardware is utterly crucial in improving the system we have described in this technical note. Once again, these concerns cannot be addressed by simply improving the precision of the underlying synchronization protocol.

\subsection{Avoiding Intermediate Cloud-Based Services}
Though the CERN SWAN service allowed direct user access to the CERNbox service, it introduced several obstacles. SWAN has no native functionality for editing JSON file types. Furthermore, the Jupyter kernel introduces further overhead, and online code refactors in the framework modules are not recognized unless the kernel is restarted. Additionally, SWAN requires a stable connection to the CERN network, which is in-demand during working hours, introducing unpredictable latency. A possible future workaround for experiments could be saving data to a local computer, rather than cloud-based service, on which the framework has been installed.

\subsection{Code Availability}
All code is available and maintained at the following public repository: \href{https://github.com/padishah115/fireball_onlineTF}{\url{https://github.com/padishah115/fireball_onlineTF}}.

\section*{Funding Acknowledgments}
This project has received funding from the European Union’s Horizon Europe Research and Innovation programme under Grant Agreement No 101057511 (EURO-LABS). Additional funding was received via the JETLAB ``Unveiling the Physics of High-Density Relativistic Pair Plasma Jets in the Laboratory''
scholarship, provided via the University of Oxford Department of Physics.

\bibliographystyle{elsarticle-num} 
\bibliography{bibliography}

\end{document}